\documentclass[prb,aps,twocolumn,amsmath,amssymb,superscriptaddress,epsf,floatfix]{revtex4}

\usepackage{bm}
\usepackage{color}
\usepackage{graphicx}

\newcommand{\bracket}[1]{\langle #1 \rangle}

\newcommand{\vect}[1]{\mbox{\boldmath $\rm #1$}}
\newcommand{\boa}{{\bf a}}
\newcommand{\bob}{{\bf b}}
\newcommand{\boc}{{\bf c}}
\newcommand{\bod}{{\bf d}}
\newcommand{\boe}{{\bf e}}
\newcommand{\bof}{{\bf f}}
\newcommand{\bog}{{\bf g}}
\newcommand{\boh}{{\bf h}}

\begin{document}

\title{Interface engineering of quantum Hall effects in digital transition-metal oxide heterostructures}

\author{Di Xiao}
\altaffiliation{e-mail: xiaod@ornl.gov; okapon@ornl.gov}
\affiliation{Materials Science and Technology Division, Oak Ridge
  National Laboratory, Oak Ridge, Tennessee 37831, USA}
  
\author{Wenguang Zhu}
\affiliation{Department of Physics and Astronomy, University of Tennessee, 
  Knoxville, Tennessee 37996, USA}
\affiliation{Materials Science and Technology Division, Oak Ridge
  National Laboratory, Oak Ridge, Tennessee 37831, USA}
  
\author{Ying Ran}
\affiliation{Department of Physics, Boston College, Chestnut Hill, MA 02467, USA}
  
\author{Naoto Nagaosa}
\affiliation{Department of Applied Physics, The University of Tokyo, 
  Hongo, Bunkyo-ku, Tokyo 113-8656, Japan}
\affiliation{Cross-Correlated Material Research Group (CMGR) and Correlated 
  Electron Research Group (CERG), RIKEN-ASI, Wako 315-0198, Japan}

\author{Satoshi Okamoto}
\altaffiliation{e-mail: xiaod@ornl.gov; okapon@ornl.gov}
\affiliation{Materials Science and Technology Division, Oak Ridge
  National Laboratory, Oak Ridge, Tennessee 37831, USA}
  
\begin{abstract}
Topological insulators are characterized by a nontrivial band topology driven by the spin-orbit coupling. 
To fully explore the fundamental science and application of topological insulators, material realization is indispensable. 
Here we predict, based on tight-binding modeling and first-principles calculations, 
that bilayers of perovskite-type transition-metal oxides grown along the [111] crystallographic axis 
are potential candidates for two-dimensional topological insulators. 
The topological band structure of these materials can be fine-tuned by changing dopant ions, substrates, and external gate voltages. 
We predict that LaAuO$_3$ bilayers have a topologically-nontrivial energy gap of about 0.15~eV, 
which is sufficiently large to realize the quantum spin-Hall effect at room temperature.
Intriguing phenomena, such as fractional quantum Hall effect, 
associated with the nearly-flat topologically-nontrivial bands found in $e_g$ systems are also discussed.
\end{abstract}

\maketitle

Since the discovery~\cite{QHE,thouless1982} of the quantum Hall effect (QHE), 
the quest for topologically ordered states of matter has become a major subject of interest in condensed matter physics.  
Haldane first proposed~\cite{haldane1988} that electrons hopping on a honeycomb lattice could realize the QHE in the absence of Landau levels, 
pointing out the possibility of nontrivial topology in simple band insulators.  
Along this direction, recent efforts have culminated in the theoretical prediction~\cite{kane2005a,bernevig2006,moore2007,fu2007} 
and subsequent experimental realization~\cite{konig2007,hsieh2008,xia2009} of the so-called topological insulators (TIs) 
in materials with strong spin-orbit coupling (SOC).  
Many interesting phenomena, including giant magneto-electric effects~\cite{qi2008} and 
the appearance of Majorana fermions~\cite{fu2008}, have been predicted.  
Once realized in real materials, these phenomena could lead to entirely new device paradigms for spintronics and quantum computing.

However, so far, the material realization of TIs has been limited to narrow band-gap semiconductors based on Hg or Bi, 
in which the electronic properties are dominated by $s$ and $p$ orbitals.  
Here, we report our theoretical investigation of topological insulating behavior in a completely different materials class%
---heterostructures of transition-metal oxides (TMOs) involving $d$ electrons.  
Our motivation is two-fold.  
First, artificial heterostructures of TMOs are becoming available owing to the recent development~\cite{izumi2001,ohtomo2002,ohtomo2004} 
in the fields of oxide superlattices and oxide electronics~\cite{mannhart2010}. 
In particular, layered structures of TMOs can be now prepared with atomic precision, 
thus offering a high degree of control over important material properties, 
such as lattice constant, carrier concentration, spin-orbit coupling, and correlation strength.  
As we show below, these advantages can be readily exploited in the design of TIs.  
Second, TMOs constitute a wide class of compounds that exhibit a variety of intriguing properties and electronic states 
associated with the electron-electron interactions, 
encompassing superconductivity, magnetism, ferroelectricity and Mott insulators.  
Combined with the TI phase, TMO heterostructures provide a very promising platform to explore various topological effects.

Our main results are summarized below.  
We first demonstrate the design principle for realizing two-dimensional TIs in bilayers of perovskite-type TMOs 
grown along the [111] crystallographic axis by using phenomenological tight-binding modeling.  
Based on this design principle and first-principles calculations, a number of candidate materials are identified.
The topological band structure of these materials can be fine-tuned by changing dopant ions, substrates, and external gate voltages, 
which will enable also the control of the topological quantum phase transition. 
In particular, we predict that LaAuO$_3$ bilayer has a topologically nontrivial energy gap about 0.15 eV, 
which is sufficiently large to realize the quantum spin-Hall effect at room temperature.
When electron-electron interaction is included, our system with topologically nontrivial band structure could have far more interesting physics.  
Here, we demonstrate this possibility by focusing on the TMO bilayers of $e_g$ systems, 
which are characterized by nearly-flat topologically nontrivial $Z_2$ bands. 
We argue that when these bands are partially filled, electron correlation could give rise to the quantum anomalous Hall effect and 
fractional quantum Hall effect.  
Our results may open new directions focusing on topological phenomena in the rapidly growing field of oxide electronics.

\begin{figure*}[tbp] 
\includegraphics[width=1.2\columnwidth,clip]{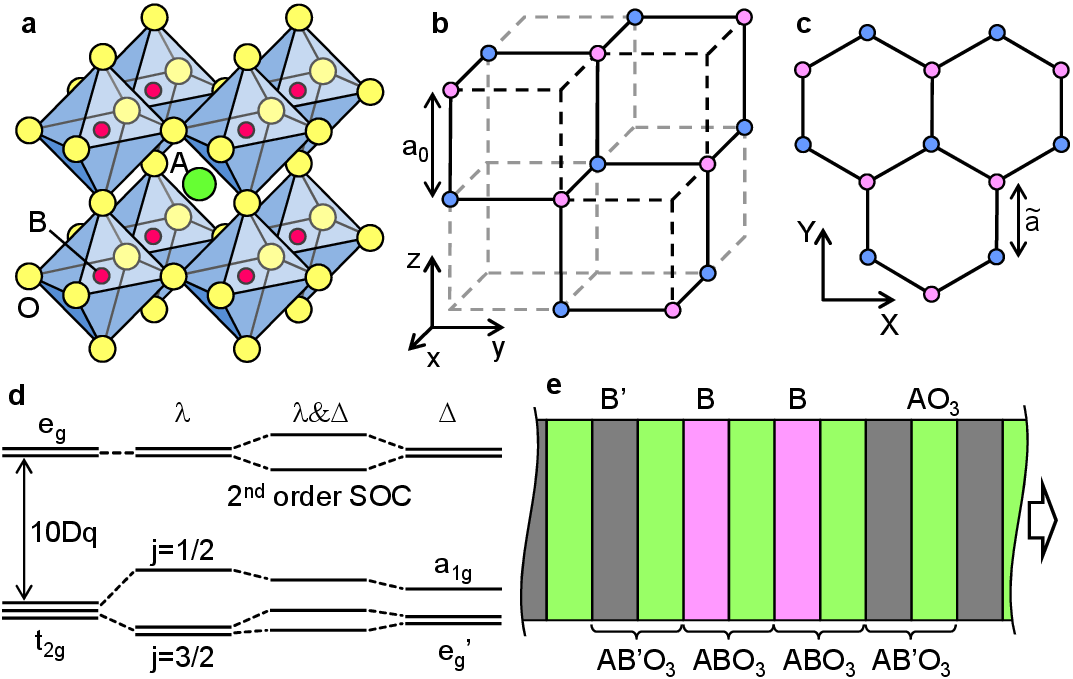} 
\caption{
{\bf Formation of the honeycomb lattice in a (111) bilayer in the cubic lattice.} 
(\boa) Perovskite structure $AB$O$_3$. 
(\bob) A (111) bilayer consisting of the top layer indicated by red circles 
and the bottom layer indicated by blue circles. The lattice constant is $a_0$. 
The bilayer shown as solid lines in (\bob) forms the honeycomb lattice when projected on the [111] plane 
with the lattice constant $\tilde a = \sqrt{2/3} a_0$ (\boc). 
The real space coordinates are labeled by $(x,y,z)$ in the original cubic lattice, while 
it is labeled by $(X,Y)$ in the [111] plane. 
(\bod) Level structure of TM $d$ orbital. 
In the cubic environment, $d$ orbitals split into $e_g$ and $t_{2g}$ manifolds. 
With the SOC, $t_{2g}$ manifold further splits into two levels characterized by the effective total angular momentum $j=1/2$ and 3/2. 
With the trigonal crystal field, $t_{2g}$ manifold splits into two levels denoted by $a_{1g}$ and $e_g'$.
With both the SOC and the trigonal field, $t_{2g}$ manifold splits into three levels and $e_g$ manifold splits into two levels, i.e., all the degeneracies are lifted except the Kramers doublets. 
(\boe) $AB$O$_3$ monolayer is grown on $A$O$_3$ terminated $AB'$O$_3$ substrate capped by $AB'$O$_3$. 
The direction of crystal growth is indicated by an arrow. }
\label{fig:model} 
\end{figure*}

\noindent\textbf{Results}

\noindent\textbf{Design principle}. To demonstrate the design principle for engineering TIs in TMO heterostructures, 
we consider perovskite-type TMOs as our prototype system.  
These compounds are very common and have the chemical formula $AB$O$_3$, where O is oxygen and $B$ is a transition-metal (TM) ion.  
The key idea is to start with a band structure that possesses ``Dirac points'' in the Brillouin zone without the SOC, 
and then examine whether an energy gap can be opened at those points with the SOC turned on.  
If an energy gap does open, combined with proper filling the resulting state could be a TI.  
In an ideal perovskite structure, the TM ions sit on a simple cubic lattice, 
with the octahedral crystalline field splitting the TM $d$ orbitals into two-fold degenerate $e_g(d_{3z^2-r^2}, d_{x^2-y^2})$ 
and three-fold degenerate $t_{2g}(d_{yz},d_{zx},d_{xy})$ levels, well separated by so-called $10Dq$ on the order of 3 eV.  
Such a lattice geometry usually does not support Dirac points.  
Instead, we consider bilayers of the perovskite structure grown in the [111] direction.  
As shown in Fig.~1, the TM ions in the (111)-bilayer are located on a honeycomb lattice consisting of two trigonal sublattices on different layers.  
This lattice geometry has three consequences: 
Firstly, 
it is well known from the study of graphene that electrons hopping on a honeycomb lattice generally give rise to Dirac points in the band structure; 
secondly, A layer potential difference can be easily created by applying a perpendicular electric field or by sandwiching the bilayer 
between two different substrates, which allows experimental control of the band topology; 
and, thirdly, the honeycomb lattice further reduces the symmetry of the crystalline field from octahedral ($O_h$) to trigonal ($C_{3v}$), 
and introduces additional level splitting of the $d$-orbitals.  
The last point turns out to be crucial for realizing the topologically insulating phase. 

We first consider the $t_{2g}$ manifold, in which the on-site SOC is active.  
In our modeling, only nearest-neighbor hopping of $d$ electrons between the TM sites via oxygen $p$ orbital is included.  
Since we are interested in the band topology, which is robust against small perturbations as long as the band gap remains open, 
our model is justified and allows us to capture the essential ingredients with minimal parameterization.  
The tight-binding Hamiltonian is given by
\begin{eqnarray}
H \!\!&=&\!\! - t \! \sum_{\vect r \vect r' \tau \tau'} \! T_{\vect r \vect r'}^{\tau\tau'} d^\dag_{\vect r \tau} d_{\vect r' \tau'}
+ \lambda \sum_{\vect r} \vect l_{\vect r} \cdot \vect s_{\vect r}  \nonumber \\
&& + \Delta \sum_{\vect r \tau\neq\tau'} d^\dag_{\vect r \tau} d_{\vect r \tau'} 
+ \frac{V}{2} \sum_{\vect r  \tau} \xi_{\vect r} d^\dag_{\vect r \tau}d_{\vect r \tau} \;,
\end{eqnarray}
where $\vect r$ and $\tau$ label the lattice sites and the $t_{2g}$ orbitals, respectively.
The first term is the hopping term represented by a single amplitude $t$ and the dimensionless structural factor $T_{\vect r \vect r'}^{\tau \tau'}$.  
The second term is the on-site SOC, which splits the $t_{2g}$ levels into a $j=1/2$ doublet with energy $\lambda$ and a $j = 3/2$ quadruplet 
with energy $-\lambda/2$.  
$\vect l_{\vect r}$ and $\vect s_{\vect r}$ are the angular momentum and spin operators. 
The third term is the trigonal crystalline field with which the $a_{1g}$-$e_g'$ splitting is given by $3 \Delta/2$.  
$V$ in the last term is the layer potential difference, and $\xi_{\vect r} = \mp 1$ when $\vect r$ is in the top or bottom layer. 
The explicit form of the Hamiltonian is presented in the Supplementary Method.  

\begin{figure*}[tbp] 
\includegraphics[width=1.2\columnwidth,clip]{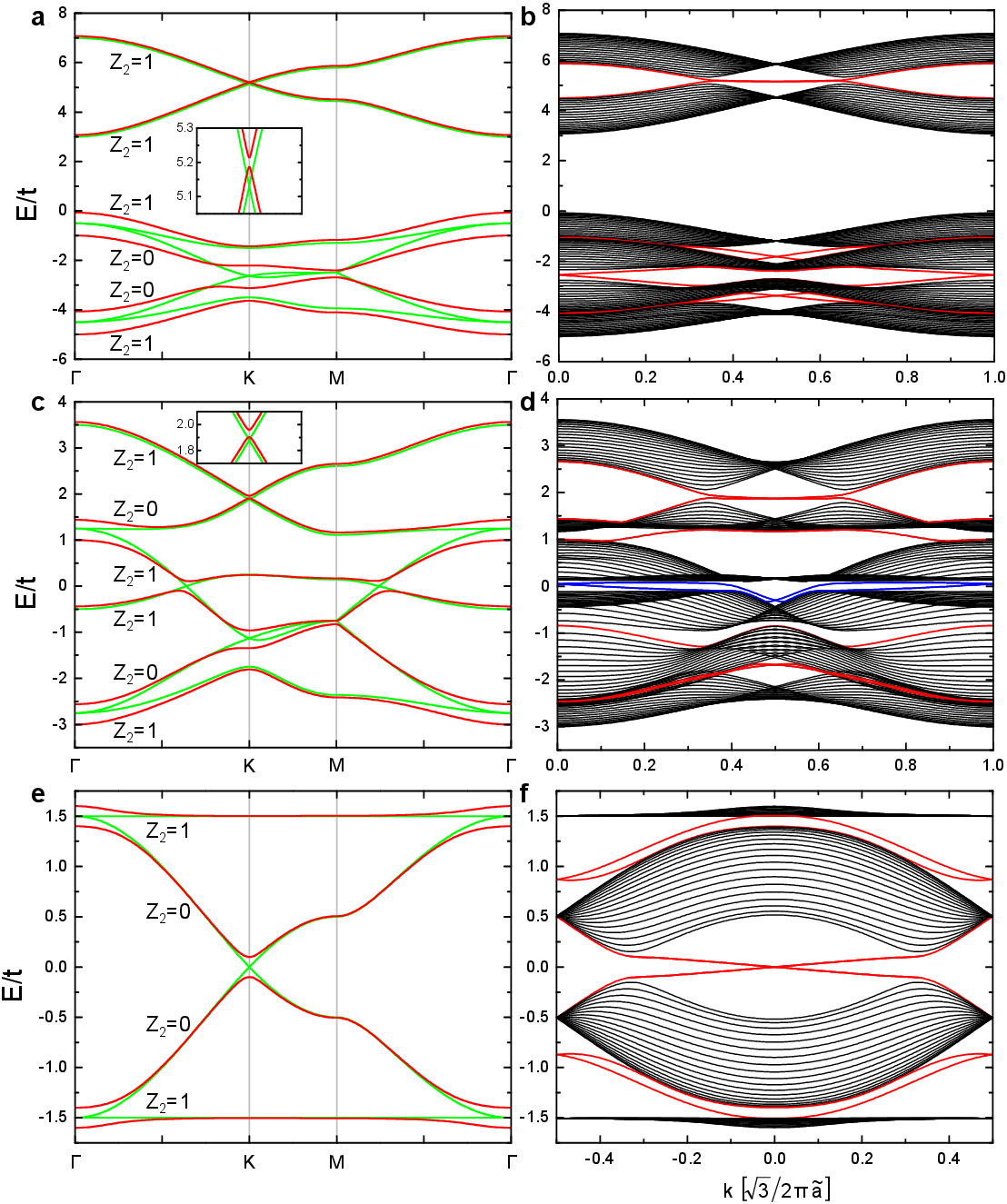}
\caption{{\bf Dispersion relations of the (111) bilayer.} 
(\boa) and (\bob) $t_{2g}$ model in the strong SOC limit. The SOC is fixed as $\lambda/t = 5$ with $\Delta/t = 1$ (red) and $\Delta=0$ (green). 
(\boc) and (\bod) $t_{2g}$ model in the weak SOC limit, $\Delta/t = 0.5$ with $\lambda/t = 1.5$ (red) and $\Delta/t = 1.5$ with $\lambda/t = 0$ (green). 
(\boe) and (\bof) $e_g$ model with $\tilde \lambda/ t= 0.2$ (red) and $\tilde \lambda / t= 0$ (green). 
Figures (\boa), (\boc), and (\boe) show the bulk dispersion relations. 
The dispersions in red correspond to the topologically nontrivial bands with the $Z_2$ invariants shown for each band. 
Sum of $Z_2$ in the occupied bands gives the $Z_2$ topological invariant for the corresponding filling. 
For example, when the lowest 5 bands of the $t_{2g}$ model are occupied by electrons in (\boa), 
$Z_2$ invariant becomes $1+0+0+1+1 \, \mbox{mod} \, 2 =1$. 
The insets in (\boa) and (\boc) show the zoom-up near the K-point. 
Figures (\bob), (\bod), and (\bof) show the dispersion relations in finite-thick zigzag ribbons 
with the periodic boundary condition along the $X$ direction and the openboundary condition along the $Y$ direction. 
Parameters are the same as in the bulk dispersions. 
Edge modes supporting the spin current are indicated by red lines. 
For the $t_{2g}$ model with the weak SOC, there appear 4 edge channels between the third and the fourth bands as shown as blue lines 
in consistent with the $Z_2$ number. 
}
\label{fig:TB} 
\end{figure*}

The large number of orbitals (6 per TM site) involved in our model give rise to a very rich behavior of the topological band structure 
in the parameter space.  
Depending on the strength of the SOC, the system falls into two different phases.  
In the strong SOC limit ($|\lambda/t| > 8/3$ when $\Delta=0$), 
bands originating from the $j = 1/2$ and $j =3/2$ orbitals are completely separated. 
The trigonal crystal field then opens up an energy gap within each manifold and a nontrivial $Z_2$ topology can be realized.  
This is similar to the results reported on Iridium compounds.~\cite{shitade2009,pesin2010}
In the weak SOC limit, bands from $j = 1/2$ and $j = 3/2$ orbitals become mixed away from the $\Gamma$ point. 
Again, we find topologically nontrivial energy gaps can be opened by $\Delta$.  
The $Z_2$ topological invariant is determined using two different methods. 
One can either evaluate it directly from the bulk band structure \cite{fu2006} or count the number of edge states.  
The calculated band structure for both cases together with the $Z_2$ index is shown in Fig.~\ref{fig:TB} a--d. 
By inspection, we find that 
$t_{2g}^1$, $t_{2g}^2$, $t_{2g}^3$ and $t_{2g}^5$ TMOs are all possible candidates for TIs in the strong SOC limit 
and $t_{2g}^2$, $t_{2g}^4$ and $t_{2g}^5$ in the weak SOC.  
The dependence of the band topology on the layer potential difference $V$ is rather interesting.  
While increasing $V$ will eventually destroy the $Z_2$ nontrivial phase, under moderate values of $V$, 
the $t_{2g}^3$ system remains TI in the strong SOC limit, and the $t_{2g}^4$ system in the weak SOC limit.

Next we consider the $e_g$ manifold.  
It is well known that the SOC is quenched within the $e_g$ manifold so it seems that the resulting band topology should be trivial.  
However, similar to graphene~\cite{min2006,yao2007} and 
some transition-metal ions~\cite{vallin1970,chen2009},
the SOC can still take place through the virtual excitation of electrons 
between $e_g$ and $t_{2g}$ levels.  According to the second order perturbation theory, the effective SOC is given by 
\begin{equation}
\widetilde H_{SO}^{\varepsilon \varepsilon'} =  \sum_{\tau \notin e_g}\frac{\bracket{\varepsilon|H_{SO}|\tau}
\bracket{\tau|H_{SO}|\varepsilon'}}{E_{e_g} - E_{\tau}} \;, 
\end{equation}
where $\varepsilon$ labels the $e_g$ orbitals.  Hence the Hamiltonian can be written 
\begin{equation}
H = - t \! \sum_{\vect r \vect r' \varepsilon \varepsilon'} \! T_{\vect r \vect r'}^{\varepsilon\varepsilon'} 
d^\dag_{\vect r \varepsilon} d_{\vect r \varepsilon'}
+ \sum_{\vect r} \widetilde H_{SO \vect r}
+ \frac{V}{2} \sum_{\vect r \varepsilon} \xi_{\vect r} d^\dag_{\vect r \varepsilon} d_{\vect r \varepsilon}, 
\end{equation}
where 
the SOC magnitude is given by $\tilde \lambda = \lambda^2/\Delta_E$, 
with $\Delta_E$ roughly being the energy difference between $e_g$ and $t_{2g}$ levels.  
(The explicit form of the Hamiltonian is presented in the Supplementary Method and Supplementary Note 1.)
We find that $\widetilde H_{SO \vect r}$ opens up an energy gap at the $\Gamma$ point and also the Dirac point located at K. 
From the inspection of the $Z_2$ topological invariant and counting the number of edge states, 
we found that $e_g^1$, $e_g^2$ and $e_g^3$ systems become TIs (See Fig.~\ref{fig:TB} e and f).
Here the trigonal crystalline field is also important---if 
all $t_{2g}$ levels are degenerate, even the second order SOC vanishes. 
A layer potential difference comparable to $\tilde \lambda$ closes a gap at the Dirac point 
turning the $e_g^2$ system into a trivial insulator. 
On the other hand, gaps at the $\Gamma$ point are stable against this perturbation. 
Instead, these gaps close when the local potential difference between $d_{3z^2-r^2}$ and $d_{x^2-y^2}$ is comparable to $\tilde \lambda$. 
Thus, the TI state and the Jahn-Teller effect \cite{jahn1937} compete in real materials with the $e_g^1$ or $e_g^3$ configuration. 

\begin{table*}[tbp]
\caption{List of candidate materials}
\label{tab:listofmaterials}
\begin{center}
\begin{tabular}{lccc} \hline 
         &   Configuration    &       Bulk                        &      Superlattice \\ \hline \hline
LaReO$_3$ &  $t_{2g}^4$   &        ---                         &      --- \\
LaRuO$_3$ &  $t_{2g}^5$   &  metallic  Ref.~\onlinecite{sugiyama1999}                   &     --- \\
SrRhO$_3$ &  $t_{2g}^5$   &   metallic  Ref.~\onlinecite{yamaura2001}                 &      Ref.~\onlinecite{europeanpatent} \\
SrIrO$_3$ &  $t_{2g}^5$   &   metallic  Refs.~\onlinecite{cao2007} \& \onlinecite{moon2008}               &      metallic Ref.~\onlinecite{sumi2005} \\
LaOsO$_3$ &  $t_{2g}^5$   &    ---                     &    --- \\
LaAgO$_3$ &  $e_{g}^2$   &  metallic (band calc.)  Ref.~\onlinecite{bacalis1988}      &      --- \\
LaAuO$_3$ &  $e_{g}^2$   &  Refs.~\onlinecite{curuswany1980} \& \onlinecite{ralle1993}                 &     --- \\
\hline
\end{tabular}
\end{center}
\end{table*}

\noindent\textbf{Materials consideration}. 
Having established that the TIs can be realized in (111)-bilayer TMO for both $t_{2g}$ and $e_g$ configurations, 
we now turn to real materials.  
We aim to realize the integer fillings established above using TM $B$ ions with the formal valence $+3$ or $+4$. 
For $B^{3+(4+)}$, we choose La (Sr) for the $A$-site element in both the target TMO and the insulating substrate $AB'$O$_3$, 
and Al (Ti) for the $B'$-site element in the insulating substrate. 
Controlling the strain effects and the layer potential difference is possible by replacing $A$ and/or $B'$ with their isovalent elements. 
It is well known that some of TMOs are insulating due to strong correlations\cite{imada1998}. 
Therefore, if the corresponding bulk system is heavily insulating, bilayering may not be useful. 
Even if the corresponding bulk system is metallic, 
the low dimensionality in (111) bilayers may drive the system into a Mott insulator\cite{yoshimatsu2010}.
%
Further, the correlation effects are expected to reduce the effective band width  
and increase the splitting between occupied levels and unoccupied levels. 
While this effect does not change the band topology in $e_g$ electron systems, this could influence the topology in 
$t_{2g}$ systems by modifying the crystal field splitting between $a_{1g}$ and $e_g'$ levels. 
In addition, in a system with an integer-number of electrons per site, 
local moments could be induced by the correlation effects resulting in the magnetic ordering. 
If the symmetry breaking by the magnetic ordering is strong, the system could become a trivial insulator. 
We do not consider such complexities by focusing on rather itinerant $4d$ and $5d$ electrons of TM ion, 
yet $t_{2g}$ electron systems are more susceptible for magnetic orderings than $e_g$ electron systems 
because of the smaller hopping intensity. 
These considerations somewhat limit the choice of TM and substrate material. 
Our candidate materials for TIs are, therefore, 
LaRe$^{3+}$O$_3$ as a $t_{2g}^4$ electron system, 
LaRu$^{3+}$O$_3$, LaOs$^{3+}$O$_3$, SrRh$^{4+}$O$_3$ and SrIr$^{4+}$O$_3$ as $t_{2g}^5$ systems, and 
LaAg$^{3+}$O$_3$ and LaAu$^{3+}$O$_3$ as $e_g^2$ electron systems. 
Most of these materials have been synthesized and their references are summarized in Table \ref{tab:listofmaterials}. 
LaReO$_3$, LaOsO$_3$ and LaAgO$_3$ have yet to be synthesized. 
According to Ref.~\onlinecite{ralle1993}, LaAuO$_3$ has CaF$_2$ structure rather than the perovskite. 
We expect this material shapes the perovskite structure by, for example, 
high-pressure synthesis and grown on a substrate with the perovskite structure. 
If properly synthesized, perovskite LaAuO$_3$ is expected to be metallic 
as LaAgO$_3$ predicted by the density functional theory calculation\cite{bacalis1988}.
We also note that growing thin films of perovskite TMOs along the [111] direction has already started\cite{Chakraverty2010a,gray2010}.
While $t_{2g}^2$ systems are also candidates for TIs, 
the TI state is hard to realize because of the band overlap.  
For $e_g^1$ and $e_g^3$ systems, additional effects 
such as longer range transfer and the Jahn-Teller effect can easily modify the dispersion relations.

\begin{figure*}[tbp] 
\includegraphics[width=1.2\columnwidth,clip]{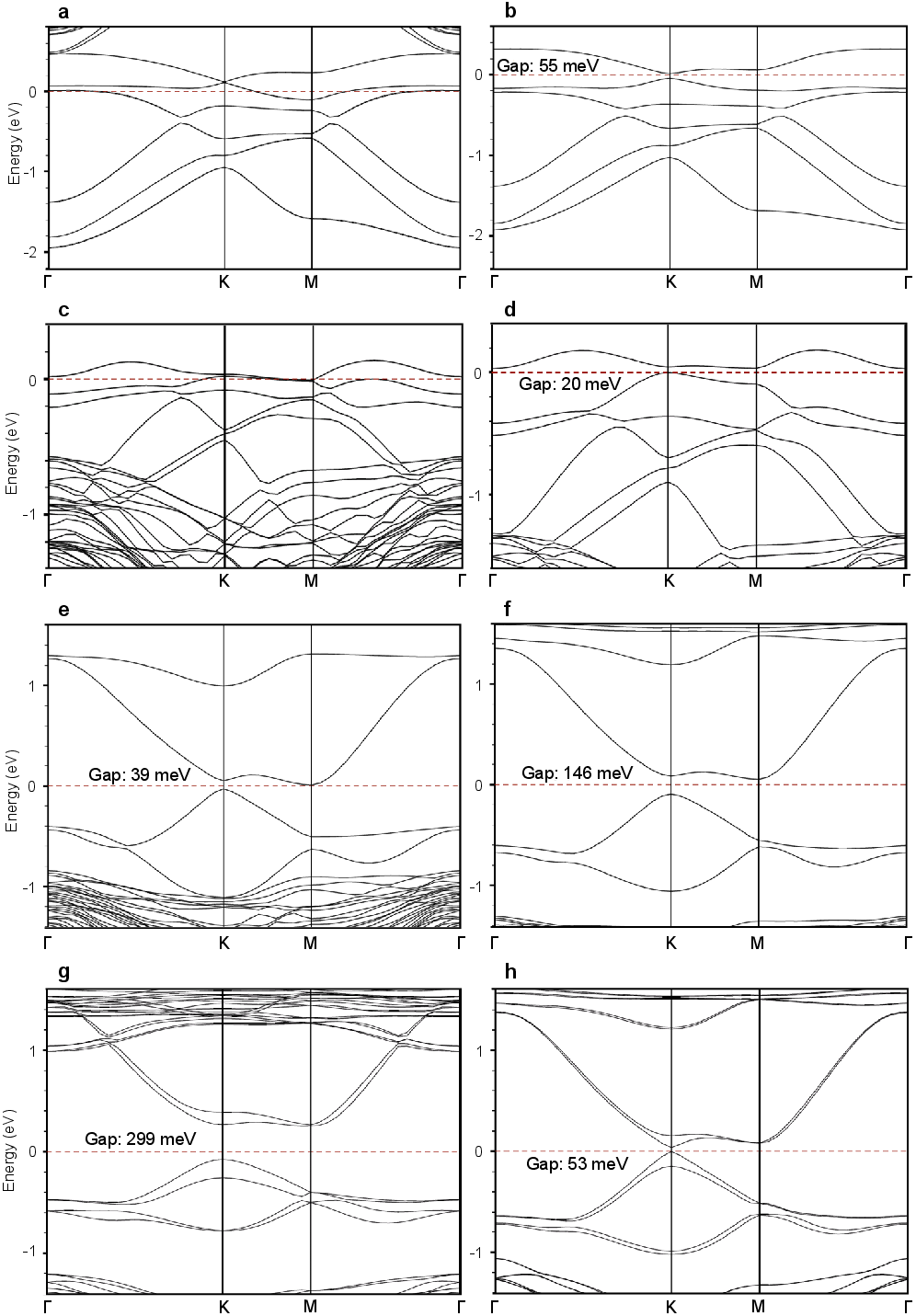}
\caption{{\bf Density functional theory results of the dispersion relations of the (111) bilayer of TMOs.} 
Symmetric bilayers: 
(\boa) LaReO$_3$, (\bob) LaOsO$_3$, (\boc) SrRhO$_3$, (\bod) SrIrO$_3$, (\boe) LaAgO$_3$, and (\bof) LaAuO$_3$. 
Bilayers shown in (\boa), (\bob), (\boe), and (\bof) are grown between LaAlO$_3$, 
while those in (\boc) and (\bod) are grown between SrTiO$_3$. 
Asymmetric bilayers of LaAuO$_3$ grown between LaAlO$_3$ and LaScO$_3$ (\bog), and between LaAlO$_3$ and YAlO$_3$ (\boh). 
The Fermi level is taken to be 0 of the vertical axis. 
Bilayers shown in (\bob), (\bod), (\boe), (\bof), and (\boh) are TIs with the band gap indicated, 
(\bog) is a trivial insulator, and others are topological metals. 
}
\label{fig:DFT} 
\end{figure*}

We first performed the density functional theory (DFT) calculations for the bilayers of $t_{2g}$ systems 
LaReO$_3$, LaRuO$_3$, LaOsO$_3$, SrRhO$_3$, SrIrO$_3$. (Details are presented in the Method section.)
Their dispersion relations are shown in Fig.~\ref{fig:DFT} a--d. 
We notice the remarkable agreement between the DFT results and the tight-binding result, Fig.~\ref{fig:TB} c, 
especially for LaReO$_3$ and LaOsO$_3$. 
For SrRhO$_3$, LaReO$_3$, and LaRuO$_3$ (not shown), the Fermi level crosses several bands. 
Thus, these systems are classified as topological metals rather than TIs.  
In LaOsO$_3$ and SrIrO$_3$, the Fermi level is located inside the gap. 
Therefore, from the analogy to the tight-binding model, (111) bilayers of LaOsO$_3$ and SrIrO$_3$ are TIs. 
From our DFT calculations, it is noted that the material dependence of the dispersion relations is rather large for $t_{2g}$ systems. 
This is because a large number of band parameters are involved in the band structure 
including the local crystalline field. 

We now move to $e_g^2$ electron systems, LaAgO$_3$ and LaAuO$_3$. 
Our DFT results for these systems are shown in Fig.~\ref{fig:DFT} e and f. 
As in the $t_{2g}$ case, the DFT reproduces the tight-binding result fairly well. 
In both undoped systems, the Fermi level is inside the gap at the K point, and 
from the analogy to the tight-binding result, these systems are TIs. 
The gap amplitude is found to be about 150~meV for LaAuO$_3$ and 40~meV for LaAgO$_3$, 
so these systems should remain TIs at room temperature. 
The band gap and topological property can be controlled by breaking the symmetry between top and bottom layers. 
As shown in Fig.~\ref{fig:DFT} g, 
the asymmetric bilayer with LaScO$_3$ has larger band gap $\sim$ 300~meV, 
and from the inspection of the symmetry of the wave function at the K point, 
this bilayer is a trivial insulator. 
While the asymmetric bilayer with YAlO$_3$ has smaller band gap $\sim$ 50~meV and remains to be a TI (Fig.~\ref{fig:DFT} h). 
Such a small band-gap TI is especially useful to control the topological property by using the gate voltage. 
Similar to LaAgO$_3$ and LaAuO$_3$, we also performed the DFT calculations for a $3d$ system LaCuO$_3$. 
We found this system develops an instability towards magnetic ordering because the itinerancy is reduced compared with $4d$ and $5d$ systems.

\begin{figure*}[tbp] 
\includegraphics[width=1.2\columnwidth,clip]{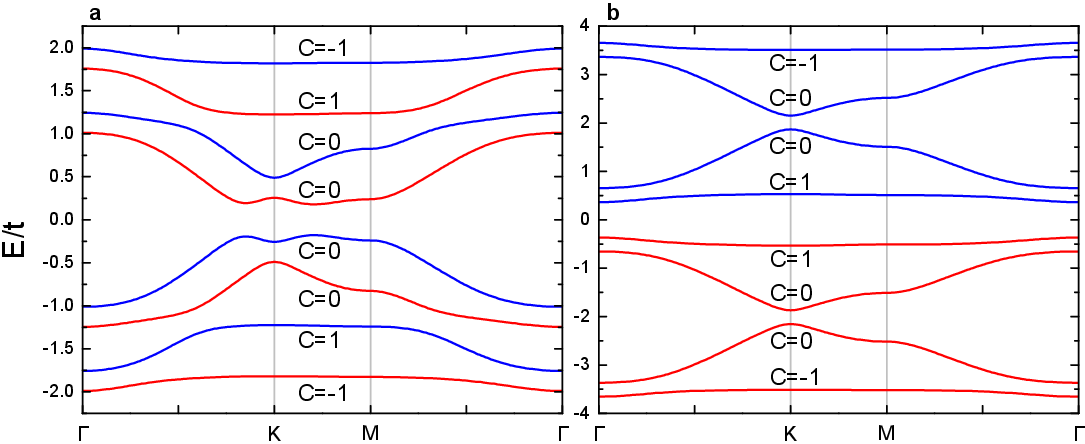}
\caption{{\bf Bulk dispersion relations of the (111) bilayer of $e_g$ model with the Zeeman splitting.} 
(\boa) Small Zeeman splitting with $B=0.3 t$, and (\bob) large Zeeman splitting with $B=2t$. 
Here we used $\tilde \lambda=0.5t$. 
Bands with the majority (minority) spin component are indicated as red (blue) lines. 
Band dependent Chern number is also indicated. 
}
\label{fig:Zeeman} 
\end{figure*}

\noindent\textbf{Nearly-flat $Z_2$ topologically nontrivial bands}. 
One of the appealing aspect of realizing nontrivial band topology in TMOs is the rich possibilities of novel phenomena 
that could emerge when electron correlation is considered.  
Here we demonstrate one of the possibilities in the $e_g$ systems with nearly-\textit{flat} $Z_2$ topologically nontrivial bands 
(see Fig.~\ref{fig:TB} e).
Obviously, when the chemical potential is tuned into the nearly-flat $Z_2$ bands, 
we are in a novel regime of quantum many-body physics. 
From the inspection of dispersion relations shown in Fig.~\ref{fig:DFT} e and f, 
the upper flat band appears to be more stable than the lower flat band. 
Tuning the chemical potential in the upper flat band corresponds to 
removing electrons from $d^{10}$ systems or adding electrons to $d^{9}$ systems. 
When such a situation is realized, kinetic energy is suppressed and physics is controlled mainly by interaction, 
whose energy scale is typically $1$--$2$~eV, much larger than the width of flat bands $W \sim 0.2$~eV. 
What would be the ground state of the system? 
Here we propose several natural candidate states. 
One very likely consequence of the short-range repulsion $U$, 
when the doping of the nearly-flat band is not too small, 
is to drive the system ferromagnetic because of Stoner's criteria $U/W\gg1$. 
While when doping is too small, Wigner crystal phase should naturally occur. 
In the following we assume that spontaneous ferromagnetic ordering occurs. 
The magnetic order should be viewed as a breaking of a discrete symmetry, due to the SOC, 
instead of the breaking of a continuous spin rotation symmetry which is not realized in a two-dimensional system at any finite temperature. 
Our discussion is not limited to spontaneous ordering, because magnetism can be also introduced externally 
by using a magnetic insulator as a substrate.

We model the effect of ferromagnetic ordering by adding a Zeeman spin splitting to the Hamiltonian: 
\begin{equation}
H_{Zeeman} = - B \sum_{\vect r \varepsilon} (d^\dag_{\vect r \varepsilon \uparrow} d_{\vect r \varepsilon \uparrow} 
- d^\dag_{\vect r \varepsilon \downarrow} d_{\vect r \varepsilon \downarrow}). 
\end{equation}
Resulting dispersion relations for $e_g$ system with two characteristic Zeeman fields are presented in Fig.~\ref{fig:Zeeman}. 
From the inspection of the Chern number, the quantum anomalous Hall (QAH) insulating state \cite{nagaosa2010} is realized 
in $e_g^{0.5}$ and $e_g^{3.5}$ systems, and also $e_g^{1}$ systems when the Zeeman field is large as in perovskite manganites. 
More interesting physics occurs when the nearly-flat ``Chern'' band is partially filled 
(See Figs.~\ref{fig:TB} e and \ref{fig:Zeeman}). 
In this case, as pointed out recently, fractional quantum Hall (FQH) states are likely to be realized
\cite{tang2010,neupert2010,sun2010,venderbos2011,wang2011,neupert2011}. 
To elaborate this possibility, we perform exact diagonalization calculation after projecting on-site repulsion 
and nearest-neighbor repulsion into the $1/3$ filled highest Chern band (i.e. at $e_g^{3.5+1/6}$). 
Indeed, signatures of a $\nu=1/3$ FQH state are observed 
(the detail of the exact diagonalization and the numerical result are presented in the Supplementary Note 2).
We find that, 
when the nearest-neighbor repulsive interaction is larger than the width of the flat band, 
the ground state degeneracy is three-fold on a torus, and the Chern number 
(excluding the integer Chern number from the filled bands) 
of the ground state wave function is $\sim 1/3$ up to finite size correction.

The aforementioned FQH effects and QAH effects are both high-temperature effects 
and fundamentally different from the quantum Hall states realized in GaAs two-dimensional electron gas in a magnetic field, 
where quasi-particle energy gap is controlled by the long-range Coulomb repulsion $e^2/(\epsilon l_B^2)$ 
($\epsilon$ is the dielectric constant and $l_B$ is the magnetic length), typically around a few Kelvin. 
In the present systems, quasi-particle energy gap is determined by the short-range repulsion $\sim$ 1--2~eV. 
This indicates that room-temperature FQH effects may be realized. 
FQH states, in particular the non-Abelian states, 
have been shown to be very useful as building blocks of a quantum computer\cite{bonderson2006,nayak2008}. 
A high temperature non-Abelian quantum Hall states in the TMO heterostructures, for example, 
at $\nu=1/2$ filling where natural candidate states are in the same universality class 
of Pfaffian states \cite{moore1991} or anti-Pfaffian states\cite{levin2007,lee2007}, 
if realized experimentally, would have strong impacts on both fundamental physics and its applications, 
including the efforts of realizing topological quantum computation.

\noindent\textbf{Discussion}

\noindent Before closing, we make few remarks on TIs in the TMO bilayers. 
The direct confirmation of the TI state is possible by measuring the conductance. 
As in Refs.~\onlinecite{bernevig2006} and \onlinecite{konig2007}, 
the conductance should be quantized as $\sigma = 2e^2/h$ per (111) bilayer in the two-terminal measurement. 
The conductance can be controlled by using the gate voltage. 
In our DFT calculations, the (111) bilayers are repeated along the [111] axis. 
With the non-zero inter-bilayer coupling, 
the helical edge channels on the surface of the sample will turn to the two Dirac fermions at 
$k_{[111]}=0$ and $1/2$ in the unit of the reciprocal lattice vector along the [111] direction. 
Thus, strictly speaking, the TI is classified as a ``weak'' TI and the backward scattering between the two Dirac fermions, 
if it exists, causes the localization. 
In order to avoid this, one needs to make the single (111) bilayer or keep the neighboring (111) bilayers far apart 
so that the inter-bilayer coupling becomes exponentially small. 
When fabricating the (111) bilayer, a small number of defects would have minor effects because the edge modes surrounding them are disconnected. 
However, as the defect density increases, two surfaces are eventually connected through the edge modes belonging to the islands. 
As a result, the backward scattering takes place. 
Another source of the localization is islands of thicker (111) TMO layers. 
The suppressed trigonal field inside such thick islands is expected to create the gapless bulk modes. 
In further thicker islands of (111) TMO layers, metallic state would be realized inside the sample. 
In addition to (111) bilayers, 
we have studied model (111) trilayers and found that the TI states are robust without 
the even-odd oscillation between TI and trivial insulator, as predicted for the bismuth thin films\cite{murakami2006}. 
Structurally, a (111) trilayer forms a so-called dice lattice, which could also bring about interesting quantum effects 
characterized by the Chern number $C = \pm2$~\cite{wangF2011}. 
Of course, if the layer structure is too thick then the bulk cubic symmetry is restored and the system is no longer a TI. 
So far, we did not mention the correlation effects and competing ground states except for the previous section. 
For limiting cases, we have performed unrestricted Hartree-Fock calculations for 
multi-orbital Hubbard models defined on (111) bilayers. 
We found that the TI states are rather robust for $e_g^2$ systems and become unstable against antiferromagnetic insulating states 
when the interaction strength is comparable to the full bandwidth 
as in the two-dimensional Hubbard model on the honeycomb lattice\cite{meng2010}. 
For $e_g^1$ or $e_g^3$ systems, 
the QAH insulating states could be generated dynamically by correlation effects without the SOC\cite{ruegg2011,yang2011}, 
yet trivial insulating states due to the Jahn-Teller effect would also be stabilized depending on 
the relative balance between the Coulomb interaction and the Jahn-Teller coupling. 
In this paper, we focused on the perovskite-type TMOs. 
Thus, our design principle for the TI state works only for the [111] plane 
because other planes such as [001] and [110] do not support a honeycomb lattice. 
However, this approach is not limited to the perovskite systems. 
For example, the [0001] plane of corundum Al$_2$O$_3$, i.e., sapphire, involves a honeycomb lattice formed by Al atoms. 
Such a system could also be utilized as the substrate material to artificially create the TI state.

\noindent\textbf{Methods} 

\noindent{\bf Tight-binding models in the real space.}
First, we consider a general multiband tight-binding (TB) model on a cubic lattice given by 
\begin{eqnarray}
H_{band} \!\!&=&\!\! 
-\sum_{\langle \vect r \vect r' \rangle \sigma} \sum_{\mu \mu'}  
\Bigl\{ t_{\vect r \vect r'}^{\mu \mu'} d^\dag_{\vect r \mu \sigma} d_{\vect r' \mu' \sigma} + h.c.\Bigr\}, 
\end{eqnarray}
where $\vect r$ labels the transition-metal sites, $\sigma$ spin, and $\mu$ orbitals. 
$t_{\vect r \vect r'}^{\mu \mu'}$ is a transfer matrix which depends on the pair of orbitals but not on the spin; 
its detail will be presented shortly.

For $t_{2g}$ electron systems, the trigonal crystal field directly couples with the local $t_{2g}$ level. 
In addition, the angular momentum is not quenched, and therefore the spin-orbit coupling (SOC) is active. 
Including these two effects, 
a TB model for $t_{2g}$ systems is written as $H_{t_{2g}}=H_{band} + H_{SO} + H_{tri}$ with 
$H_{SO}$ and $H_{tri}$ given by the second and the third terms of Eq. (1), respectively. 
The explicit form of $H_{SO}$ for the $t_{2g}$-alone model is given by 
\begin{equation}
H_{SO} = \lambda \sum_{\vect r} \vect l_{\vect r} \cdot \vect s_{\vect r} =
\frac{\lambda}{2} \sum_{\vect r \sigma \sigma'} \sum_{\tau \tau' \tau''}  
i \varepsilon_{\tau \tau' \tau''} d^\dag_{\vect r \tau \sigma} 
\sigma^{\tau''}_{\sigma \sigma'} 
d_{\vect r \tau' \sigma}, 
\label{eq:Hso}
\end{equation}
with the use of the following convention for the orbital index: 
$|a \rangle=|d_{yz} \rangle$, $|b \rangle=|d_{zx} \rangle$, and $|c \rangle=|d_{xy} \rangle$.
$\sigma^\tau$ 
with $\tau=a,b,c$ is the Pauli matrix, 
and $\varepsilon_{\tau \tau' \tau''}$ is the Levi-Civita antisymmetric tensor. 
%
%

The dependence of transfer matrices on the orbital and direction 
is given by the Slater-Koster formula \cite{slater54} 
as follows: 
\begin{eqnarray}
t_{\vect r, \vect r \pm \hat{\vect y} (\hat{\vect z}) }^{a a} \!\!&=&\!\! 
t_{\vect r, \vect r \pm \hat{\vect z} (\hat{\vect x}) }^{b b} = 
t_{\vect r, \vect r \pm \hat{\vect x} (\hat{\vect y}) }^{c c} = t_\pi, \\
t_{\vect r, \vect r \pm \hat{\vect x} }^{a a} \!\!&=&\!\! 
t_{\vect r, \vect r \pm \hat{\vect y} }^{b b} = t_{\vect r, \vect r \pm \hat{\vect z} }^{c c} = t_{\delta'}, 
\end{eqnarray}
for the nearest-neighbor (NN) hopping 
and 
\begin{eqnarray}
t_{\vect r, \vect r \pm \hat{\vect y} \pm \hat{\vect z}}^{a a} \!\!&=&\!\! 
t_{\vect r, \vect r \pm \hat{\vect z} \pm \hat{\vect x}}^{b b} = 
t_{\vect r, \vect r \pm \hat{\vect y} \pm \hat{\vect z}}^{c c} = t_{\sigma''} \\
t_{\vect r, \vect r \pm (\hat{\vect x} + \hat{\vect y})}^{a b} \!\!&=&\!\!  
t_{\vect r, \vect r \pm (\hat{\vect y} + \hat{\vect z})}^{b c} = 
t_{\vect r, \vect r \pm (\hat{\vect z} + \hat{\vect x})}^{c a} = t_{\pi'}, \\
t_{\vect r, \vect r \pm (\hat{\vect x} - \hat{\vect y})}^{a b} \!\!&=&\!\!  
t_{\vect r, \vect r \pm (\hat{\vect y} - \hat{\vect z})}^{b c} = 
t_{\vect r, \vect r \pm (\hat{\vect z} - \hat{\vect x})}^{c a} = - t_{\pi'}, \hspace{1em}
\end{eqnarray}
for the second-neighbor (SN) hopping. 
Here, $\hat{\vect x}$, $\hat{\vect y}$, and $\hat{\vect z}$ are the unit vector along the x-, y-, and z-direction, respectively. 
Although it is via weak $\pi$ hybridization $t_{pd}^\pi$ between a transition-metal ion and an oxygen ion, 
the NN hopping $t_\pi \propto (t_{pd}^\pi)^2/\Delta_{pd}$ is the largest parameter in this model, thus, taken as the unit of energy $t$. 
$\Delta_{pd}$ is the level difference between TM $d$ orbitals and oxygen $p$ orbitals. 
The ratio between $t_{\vect r \vect r'}^{\tau \tau'}$ and $t_\pi$ is the dimensionless parameter 
$T_{\vect r \vect r'}^{\tau \tau'} = t_{\vect r \vect r'}^{\tau \tau'}/t_\pi$. 
$t_{\delta'}$ is also the NN hopping, but it is via weak direct overlap and, therefore, is expected to be small. 
$t_{\sigma''}$ and $t_{\pi'}$ are the SN hoppings due to the higher-order processes involving the transfer between two oxygen ions 
as $t_{\sigma'',\pi'} \propto (t_{pd}^\pi)^2 t_{pp}^{\sigma, \pi}/\Delta_{pd}^2$. 
Since $t_{\sigma'' (\pi')}$ involves relatively strong (weak) $\sigma (\pi)$ hybridization between two oxygen ions $t_{pp}^{\sigma (\pi)}$, 
we expect $|t_{\sigma''}| > |t_{\pi'}|$. 
Typical transfer intensities are shown in Supplementary Figure~S1 (a). 

For $e_g$ electron systems, 
linear coupling with the trigonal crystal field is absent. 
Therefore, the $C_3$ lattice symmetry of a (111) bilayer does not influence the on-site $e_g$ level. 
On the other hand, $e_g$ degeneracy can be lifted by the distortion of an O$_6$ cage surrounding a transition-metal ion, 
i.e., the Jahn-Teller effect. 
Focusing on the metallic regime, we neglect this effect. 
The angular momentum is quenched unless the coupling between $e_g$ and $t_{2g}$ orbitals are considered. 
We also neglect this effect at the moment but reconsider it later. 
Thus, for the $e_g$-alone model, 
$H_{e_g} = H_{band}$. 
The dependence of transfer matrices on the orbital and direction 
is again given by the Slater-Koster formula\cite{slater54}.
For the NN hopping, we have 
\begin{eqnarray}
t_{\vect r, \vect r \pm \hat{\vect z}}^{\varepsilon \varepsilon'} \!\!&=&\!\! 
\left[
\begin{matrix}
t_{\vect r, \vect r \pm \hat{\vect z}}^{\alpha \alpha} & t_{\vect r, \vect r \pm \hat{\vect z}}^{\alpha \beta} \\
t_{\vect r, \vect r \pm \hat{\vect z}}^{\beta \alpha} & t_{\vect r, \vect r \pm \hat{\vect z}}^{\beta \beta}
\end{matrix}
\right]
=
\biggl[
\begin{matrix}
t_\sigma & 0 \\
0 & t_\delta
\end{matrix}
\biggr], \\
t_{\vect r, \vect r \pm \hat{\vect x}}^{\varepsilon \varepsilon'} \!\!&=&\!\! 
\frac{1}{4}
\left[
\begin{matrix}
t_\sigma + 3 t_\delta &  - \sqrt{3} (t_\sigma - t_\delta) \\
- \sqrt{3} (t_\sigma - t_\delta) & 3 t_\sigma + t_\delta
\end{matrix}
\right], \\
t_{\vect r, \vect r \pm \hat{\vect y}}^{\varepsilon \varepsilon'} \!\!&=&\!\! 
\frac{1}{4}
\left[
\begin{matrix}
t_\sigma + 3 t_\delta &  \sqrt{3} (t_\sigma - t_\delta) \\
\sqrt{3} (t_\sigma - t_\delta) & 3 t_\sigma + t_\delta
\end{matrix}
\right], 
\end{eqnarray}
and for the SN hopping 
\begin{eqnarray}
t_{\vect r, \vect r \pm \hat{\vect x} \pm \hat{\vect y}}^{\varepsilon \varepsilon'} \!\!&=&\!\! 
\frac{1}{2}t_{\sigma'}
\left[
\begin{matrix}
1 &  0 \\
0 & - 3
\end{matrix}
\right], \\
t_{\vect r, \vect r \pm \hat{\vect z} \pm \hat{\vect x}}^{\varepsilon \varepsilon'} \!\!&=&\!\! 
-\frac{1}{2}t_{\sigma'}
\left[
\begin{matrix}
2 &  - \sqrt{3} \\
- \sqrt{3} & 0
\end{matrix}
\right], \\
t_{\vect r, \vect r \pm \hat{\vect z} \pm \hat{\vect y}}^{\varepsilon \varepsilon'} \!\!&=&\!\! 
-\frac{1}{2}t_{\sigma'}
\left[
\begin{matrix}
2 & \sqrt{3} \\
\sqrt{3} & 0
\end{matrix}
\right]. 
\end{eqnarray}
Here, $\varepsilon (= \alpha, \beta)$ labels the $e_g$ orbitals as 
$|\alpha \rangle=|d_{3z^2-r^2} \rangle$ and $|\beta \rangle=|d_{x^2-y^2} \rangle$. 
For $e_g$ electron systems, 
the NN hopping $t_\sigma \propto (t_{pd}^\sigma)^2/\Delta_{pd}$ is via strong $\sigma$ hybridization 
between a transition-metal ion and an oxygen ion $t_{pd}^\sigma$ and, therefore, largest. 
This hopping integral is taken as the unit of energy $t$. 
Again, the dimensionless parameter $T_{\vect r \vect r'}^{\varepsilon \varepsilon'}$ is defined by the ratio between 
$t_{\vect r \vect r'}^{\varepsilon \varepsilon'}$ and $t_\sigma$ as 
$T_{\vect r \vect r'}^{\varepsilon \varepsilon'}= t_{\vect r \vect r'}^{\varepsilon \varepsilon'}/t_\sigma$. 
The NN hopping $t_\delta$ is due mainly to the direct overlap between two TM ions and, therefore, expected to be small 
$|t_\delta| \ll |t_\sigma|$ as $t_{\delta'}$ in the $t_{2g}$ orbital model. 
$t_{\sigma'}$ is the SN hopping due to the higher-order processes involving the transfer between two oxygen ions as 
$t_{\sigma'} \propto (t_{pd}^\sigma)^2 t_{pp}^\sigma / \Delta_{pd}^2$. 
Typical transfer intensities are shown in Supplementary Figure~S1 (b). 

\vspace{1em}
\noindent{\bf Tight-binding models on the (111) bilayer.}
By constraining the atomic coordinate $\vect r$ within the (111) bilayer $\vect R = (X,Y)$, 
it is straightforward to derive the TB Hamiltonian as a function of two-dimensional momentum $\vect k = (k_x, k_y)$. 
We use the convention in which the projection of the nearest-neighbor bond into the [111] plane 
$\tilde a$ is taken as the unit of the length scale. 
This is a factor $\sqrt{2/3}$ smaller than the lattice constant of the cubic perovskite, and 
the size of the new unit cell is $3 \sqrt{3} \tilde a^2/2$. 
Taking the primitive lattice vectors 
as $\vect a_1 = (\sqrt{3} \tilde a,0)$ and $\vect a_2 = (\sqrt{3} \tilde a /2, 3 \tilde a /2)$, 
the first Brillouin zone is a hexagon with 6 corners located at 
$\vect k = (\pm 4 \pi/3 \sqrt{3} \tilde a, 0), (\pm 2 \pi/3 \sqrt{3} \tilde a, \pm 2 \pi/3 \tilde a)$.

It is straightforward to derive the Hamiltonian matrix in the (111) bilayer. 
For the $t_{2g}$ orbital model, 
we obtain 
\begin{widetext}
\begin{equation}
H_{band} + H_{tri} = \sum_{\vect k} 
\left[
\begin{matrix}
d_{1 a \vect k}^\dag \\ d_{1 b \vect k}^\dag \\ d_{1 c \vect k}^\dag \\ 
d_{2 a \vect k}^\dag \\ d_{2 b \vect k}^\dag \\ d_{2 c \vect k}^\dag 
\end{matrix}
\right]^T 
\left[
\begin{array}{ccc|ccc}
-\frac{V}{2} + \tilde \varepsilon_{a \vect k} & \gamma_{c \vect k} & \gamma_{b \vect k} & 
\varepsilon_{a \vect k} &  &  \\
\gamma_{c \vect k} & 
-\frac{V}{2} + \tilde \varepsilon_{b \vect k} & \gamma_{a \vect k} &
 & \varepsilon_{b \vect k} & \\
\gamma_{b \vect k} & \gamma_{a \vect k} & 
-\frac{V}{2} + \tilde \varepsilon_{c \vect k} & 
 & & \varepsilon_{c \vect k} \\ \hline
\varepsilon_{a \vect k}^* & & &
\frac{V}{2} + \tilde \varepsilon_{a \vect k} & \gamma_{c \vect k} & \gamma_{b \vect k}  \\
 & \varepsilon_{b \vect k}^* & & 
\gamma_{c \vect k} & 
\frac{V}{2} + \tilde \varepsilon_{b \vect k} & \gamma_{a \vect k} \\
 & & \varepsilon_{c \vect k}^* & 
\gamma_{b \vect k} & \gamma_{a \vect k} & 
\frac{V}{2} + \tilde \varepsilon_{c \vect k} 
\end{array}
\right] 
\left[
\begin{array}{c}
d_{1 a \vect k} \\ d_{1 b \vect k} \\ d_{1 c \vect k} \\ 
d_{2 a \vect k} \\ d_{2 b \vect k} \\ d_{2 c \vect k} 
\end{array}
\right], 
\end{equation}
\end{widetext}
with
%
\begin{eqnarray}
\varepsilon_{a \vect k} \!\!&=&\!\! -t_\pi \Bigl\{ \! 1 + e^{i \bigl(\frac{\sqrt{3}}{2} k_x - \frac{3}{2} k_y \bigr)} \! \Bigr\} 
- t_{\delta'} e^{-i \bigl(\frac{\sqrt{3}}{2} k_x + \frac{3}{2} k_y\bigr)} \!, 
\end{eqnarray}
\begin{eqnarray}
\varepsilon_{b \vect k} \!\!&=&\!\! -t_\pi \Bigl\{ \! 1 + e^{-i \bigl(\frac{\sqrt{3}}{2} k_x + \frac{3}{2} k_y \bigr)} \! \Bigr\} 
- t_{\delta'} e^{i \bigl(\frac{\sqrt{3}}{2} k_x - \frac{3}{2} k_y \bigr)} \!, 
\end{eqnarray}
\begin{eqnarray}
\varepsilon_{c \vect k} \!\!&=&\!\! -t_{\delta'} 
- 2 t_\pi \cos \Bigl( \frac{\sqrt{3}}{2} k_x \Bigr) e^{- i \frac{3}{2} k_y}, \\
\gamma_{a \vect k} \!\!&=&\!\! \frac{1}{2}\Delta + 2 t_{\pi'} \cos \Bigl( - \frac{\sqrt{3}}{2} k_x + \frac{3}{2} k_y \Bigr), \\ 
\gamma_{b \vect k} \!\!&=&\!\! \frac{1}{2}\Delta + 2 t_{\pi'} \cos \Bigl( \frac{\sqrt{3}}{2} k_x + \frac{3}{2} k_y \Bigr), \\
\gamma_{c \vect k} \!\!&=&\!\! \frac{1}{2}\Delta + 2 t_{\pi'} \cos \sqrt{3} k_x , \\
\tilde \varepsilon_{a \vect k} \!\!&=&\!\! - 2 t_{\sigma''} \cos \Bigl( -\frac{\sqrt{3}}{2} k_x + \frac{3}{2} k_y \Bigr), \\
\tilde \varepsilon_{b \vect k} \!\!&=&\!\! - 2 t_{\sigma''} \cos \Bigl( \frac{\sqrt{3}}{2} k_x + \frac{3}{2} k_y \Bigr), \\
\tilde \varepsilon_{c \vect k} \!\!&=&\!\! - 2 t_{\sigma''} \cos \sqrt{3} k_x .
\end{eqnarray}
%
%
%
%
Here, the spin indices are suppressed for simplicity. 
$\mp V/2$ is the sublattice-dependent potential which breaks the symmetry between the top (labeled 1) and bottom (labeled 2) layers.

\begin{widetext}
For the $e_g$ orbital model, we obtain 
\begin{eqnarray}
&&\hspace{-3em}H_{band} = \sum_{\vect k} 
\left[
\begin{matrix}
d_{1 \alpha \vect k}^\dag \\ d_{1 \beta \vect k}^\dag \\ d_{2 \alpha \vect k}^\dag \\ d_{2 \beta \vect k}^\dag
\end{matrix}
\right]^T \!\! 
%
\left[
\begin{matrix}
-\frac{V}{2}
+ \tilde \varepsilon_{\alpha \vect k} & \tilde \varepsilon_{\alpha \beta \vect k} & 
\varepsilon_{\alpha \vect k} & \varepsilon_{\alpha \beta \vect k} \\
\tilde \varepsilon_{\alpha \beta \vect k} & 
-\frac{V}{2}
+ \tilde \varepsilon_{\beta \vect k} & 
\varepsilon_{\alpha \beta \vect k} & \varepsilon_{\beta \vect k} \\
\varepsilon_{\alpha \vect k}^* & \varepsilon_{\alpha \beta \vect k}^* & 
\frac{V}{2}
+ \tilde \varepsilon_{\alpha \vect k} & \tilde \varepsilon_{\alpha \beta \vect k} \\
\varepsilon_{\alpha \beta \vect k}^* & \varepsilon_{\beta \vect k}^* & 
\tilde \varepsilon_{\alpha \beta \vect k} & 
\frac{V}{2}
+ \tilde \varepsilon_{\beta \vect k} 
\end{matrix}
\right]
\!\!\!
\left[
\begin{matrix}
d_{1 \alpha \vect k} \\ d_{1 \beta \vect k} \\ d_{2 \alpha \vect k} \\ d_{2 \beta \vect k}
\end{matrix}
\right] \!\! ,
\label{eq:TBeg}
\end{eqnarray}
\end{widetext}
with 
\begin{eqnarray}
%
%
\varepsilon_{\alpha \vect k} \!\!&=&\!\! 
-t_\sigma - \frac{1}{2}(t_\sigma +3 t_\delta) \cos \biggl(\frac{\sqrt{3}}{2} k_x \biggr) e^{-i \frac{3}{2}k_y}, \\
\varepsilon_{\beta \vect k} \!\!&=&\!\! -t_\delta - \frac{1}{2}(3 t_\sigma + t_\delta) 
\cos \biggl(\frac{\sqrt{3}}{2} k_x \biggr) e^{-i \frac{3}{2}k_y},  \\
\varepsilon_{\alpha \beta \vect k} \!\!&=&\!\! - \frac{\sqrt{3}}{2} (t_\sigma - t_\delta) 
i \sin \biggl(\frac{\sqrt{3}}{2} k_x \biggr) e^{-i \frac{3}{2}k_y}, \\
\tilde \varepsilon_{\alpha \vect k} \!\!&=&\!\! 
t_{\sigma'} \biggl\{ 4  \cos \biggl(\frac{\sqrt{3}}{2} k_x \biggr) \cos \biggl(\frac{3}{2} k_y \biggr) 
- \cos \sqrt{3} k_x \biggr\} \!, \hspace{2.5em}\\
\tilde \varepsilon_{\beta \vect k} \!\!&=&\!\! 3 t_{\sigma'} \cos \sqrt{3} k_x, \\
\tilde \varepsilon_{\alpha \beta \vect k} \!\!&=&\!\! 
2 \sqrt{3} t_{\sigma'} \sin \biggl(\frac{\sqrt{3}}{2} k_x \biggr) \sin \biggl(\frac{3}{2} k_y \biggr). 
\end{eqnarray}

\noindent{\bf DFT calculations.} 
Density functional calculations were carried out using the projector augmented wave (PAW) method \cite{kresse1999} 
with the generalized gradient approximation (GGA) in the parametrization of Perdew, Burke, and Enzerhof (PBE) \cite{perdew1996} 
for exchange-correlation as implemented in the Vienna Ab Initio Simulation Package (VASP) \cite{kresse1996}. 
The default plane-wave energy cutoff for O, 400.0 eV, was consistently used in all the calculations. 
The optimized crystal parameter is 3.81~\AA \, for bulk LaAlO$_3$ and 3.95~\AA \, for bulk SrTiO$_3$. 
These values are in consistent with the experimental values of 3.79~\AA \, (Ref~\onlinecite{berkstresser1991}) and 
3.91~\AA (Ref.~\onlinecite{hellwege81}). 
The transition-metal bilayer structures were simulated by a supercell consisting of 12 $A$O$_3$ and 12 $B$ layers along the [111] direction 
with ($A,B$)=(La,Al) or (Sr,Ti)
with two adjacent $B$ layers replaced by transition-metal ions. 
In the (111) plane, the supercell contains a $1 \times 1$ unit cell. 
A $6 \times 6 \times 1$ special $\vect k$-point mesh including the $\Gamma$ point (0,0,0) was used for integration over the Brillouin zone. 
Optimized atomic structures were achieved when forces on all the atoms were less than 0.01 eV/\AA.

\noindent\textbf{Acknowledgements}

\noindent
The authors thank H. Christen, 
K. Yamaura, Wanxiang Feng, and A. R{\"u}egg  for their stimulating discussions.  
Work by W.Z. and S.O. was supported by the U.S. Department of Energy, Office of Basic Energy Sciences,
Materials Sciences and Engineering Division. 
D.X. was supported by the Laboratory Directed Research and Development Program of ORNL.  
N.N. acknowledges support from MEXT Grant-in-Aid for Scientific Research (19048008, 19048015, 21244053).
Computational support was provided by NERSC of U.S. Department of Energy.

\noindent\textbf{Author contributions}

\noindent
S.O. conceived the idea for the (111) bilayer and constructed the theoretical models. 
Model calculations were performed by D.X. and S.O. (non-interacting models), and Y.R. (interacting models).
The density functional theory calculations were performed by W.Z. 
N.N. provided further theoretical inputs. 
S.O. and D.X. were responsible for overall project direction, planning and management. 

\noindent\textbf{Competing financial interests} The authors declare no competing financial interests.


%
\onecolumngrid
%
%
%
%
%

\renewcommand{\thetable}{S\Roman{table}}
\renewcommand{\thefigure}{S\arabic{figure}}
\renewcommand{\thesubsection}{Supplementary Note \arabic{subsection}}
\renewcommand{\theequation}{S \arabic{equation}}

\setcounter{secnumdepth}{3}




\newpage

\begin{center}
{\bf Supplementary Figures}
\end{center}

\begin{center}
\includegraphics[width=0.4\columnwidth,clip]{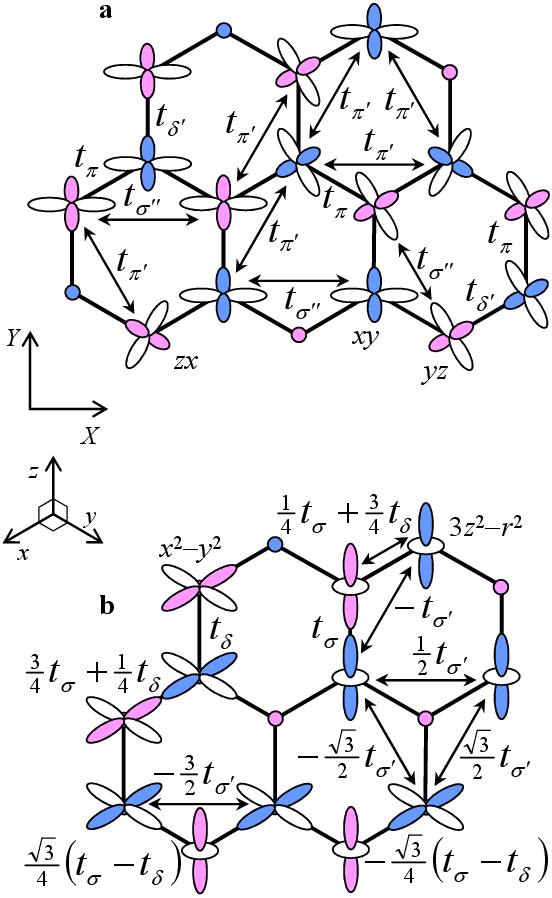} 
\begin{minipage}{0.7\columnwidth}
\vspace{0.2in}
\noindent{\textbf{Supplementary Figure S1.} 
Orbital dependent transfer integral on a (111) bilayer for 
({\bf a}) $t_{2g}$ model and ({\bf b}) $e_g$ model. 
Wave functions and lattice sites on the top layer are indicated by red and those on the bottom layer are indicated by blue.} 
\end{minipage}
\end{center}


\vspace*{0.5in}
\begin{center}
\includegraphics[scale=0.5]{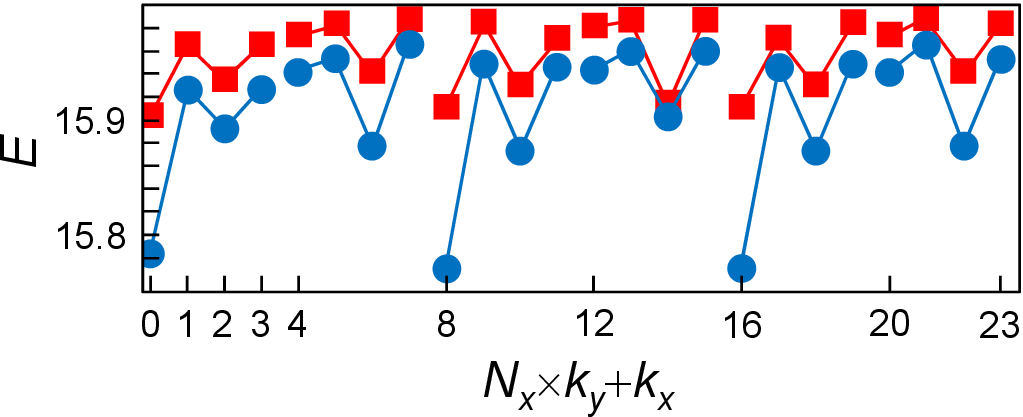} 
\begin{minipage}{0.7\columnwidth}
\vspace{0.2in}
\noindent{\textbf{Supplementary Figure S2.} 
Energy spectrum of the effective model Eq.~(\ref{eq:Heff}) defined on a $4\times6$ cluster with 8 electrons. 
Nearly degenerate GSM and the large gap between GSM and excited states 
indicate the formation of the fractional quantum Hall states.}
\end{minipage}
\end{center}

\newpage

\vspace{2em}
\begin{center}
{\bf Supplementary Notes}
\end{center}

\renewcommand{\theequation}{S1.\arabic{equation}}

\setcounter{secnumdepth}{3}

\begin{center}
{\bf Supplementary Note 1. Effective spin-orbit coupling for $\mbox{\boldmath $e_g$}$ system in the (111) bilayer}
\end{center}

\setcounter{equation}{0}

For an $e_g$ model in the cubic symmetry, the SOC is absent because the angular momentum is quenched. 
However in (111) bilayers, the local symmetry is lowered to $C_3$ and, therefore, 
the virtual coupling with $t_{2g}$ orbital can induce the effective SOC between the two $e_g$ orbitals. 

To see this effect, we consider the $a_{1g}$ orbital given by the linear combination of three $t_{2g}$ orbitals. 
This state is expected to be highest in energy among three orthogonal orbitals originating from $t_{2g}$ triplets. 
Two $e_g$ orbitals are expressed as $|\alpha \rangle = |d_{3z^2-r^2} \rangle = |m=0 \rangle$ and 
$|\beta \rangle = |d_{x^2-y^2} \rangle = \frac{1}{\sqrt{2}} \{|2 \rangle + |- 2 \rangle \}$, 
where the azimuthal quantum number $l=2$ is suppressed for simplicity. 
Here, the spin quantization axis is taken to be $z$ in the original cubic lattice. 
The wave function for $a_{1g}$ is expressed as 
\begin{eqnarray}
|a_{1g} \rangle \!\!&=&\!\! \frac{1}{\sqrt{3}} \bigl\{|d_{yz}\rangle+|d_{zx}\rangle+|d_{xy}\rangle \bigr\} \nonumber \\
\!\!&=&\!\!
\frac{1}{\sqrt{6}} \bigl\{-(1-i)|1\rangle + (1+i)|-1\rangle -i \bigl( |2\rangle - |-2\rangle \bigr) \bigr\}. \nonumber\\
\end{eqnarray}
Using the atomic SOC, $H_{SO}=\lambda \vect l \cdot \vect s = \lambda \{ l_z s_z + \frac{1}{2} (l_+ s_- + l_- s_+)\}$, 
as a perturbation, and the level separation $\Delta_E$ between $e_g$ and $a_{1g}$, 
we obtain the effective SOC for the $e_g$-alone model as 
\begin{eqnarray}
\sum_{\vect r} \widetilde H_{SO \vect r} =  \frac{\tilde \lambda}{\sqrt{12}} \sum_{\vect r} 
\left[
\begin{matrix}
d_{\vect r \alpha \uparrow}^\dag \\
d_{\vect r \beta \uparrow}^\dag \\
d_{\vect r \alpha \downarrow}^\dag \\
d_{\vect r \beta \downarrow}^\dag \\
\end{matrix}
\right]^T
\left[
\begin{array}{cc|cc}
   & i & & 1+i \\
 -i & & -1-i \\ \hline
   & -1+i & & -i \\
1-i & & i & 
\end{array}
\right]
\left[
\begin{matrix}
d_{\vect r \alpha \uparrow} \\
d_{\vect r \beta \uparrow} \\
d_{\vect r \alpha \downarrow} \\
d_{\vect r \beta \downarrow} \\
\end{matrix}
\right].
\label{eq:SOeg}
\end{eqnarray}
%
The diagonal term $\tilde \lambda/2 \hat I_6$ is absorbed in the chemical potential. 
The effective SOC constant is given by $\tilde \lambda = \lambda^2/\Delta_E$. 
Therefore, when the level separation between $e_g$ and $a_{1g}$ orbitals is small, 
the effective SOC $\tilde \lambda$ could become large. 
Other contributions including the excitation to $e_g'$ orbitals are expected to normalize $\tilde \lambda$. 
In particular, when $a_{1g}$ and $e_g'$ orbitals are degenerate, i.e., $\Delta=0$, 
off-diagonal components of Eq.~(\ref{eq:SOeg}) disappear. 
By changing the spin quantization axis along the [111] direction, 
the effective SOC is conveniently written as 
\begin{equation}
\sum_{\vect r} \widetilde H_{SO \vect r} =  - \frac{\tilde \lambda}{2} \sum_{\vect r}
d^\dag_{\vect r \varepsilon \sigma} \tau^y_{\varepsilon \varepsilon'} \sigma^z_{\sigma \sigma'} d_{\vect r \varepsilon' \sigma'},
\end{equation}
where $\tau^y$ is the Pauli matrix acting in the orbital space. 

\renewcommand{\theequation}{S2.\arabic{equation}}

\begin{center}
{\bf Supplementary Note 2. $\mbox{\boldmath $\nu=1/3$}$ Fractional Quantum Hall state on $\mbox{\boldmath $e_g$}$ flat band}
\end{center}
\setcounter{equation}{0}

The tight-binding model that we are studying is given by Eqs.~(28) and (\ref{eq:SOeg}). 
We choose $\tilde \lambda=0.5 t,B=0.3t$, and the 8-bands are shown in Fig. 4 a.

Now we focus on the highest band -- the 8th band (it differs from the
lowest band by a particle hole transformation), and consider the electron interactions $H_{I}$. 
For $H_{I}$, we use the following interaction:
\begin{eqnarray}
H_{I}=U\sum_{\vect r \varepsilon}n_{\vect r \varepsilon \uparrow}n_{\vect r \varepsilon \downarrow}+
U'\sum_{\vect r \varepsilon>\varepsilon'}n_{\vect r \varepsilon}n_{\vect r \varepsilon'}
+V\sum_{\langle \vect r \vect r' \rangle}n_{\vect r}n_{\vect r'}.
\end{eqnarray}
Here $\vect r$ labels site, $\varepsilon$ labels $e_g$ orbitals, and $\sigma=\uparrow,\downarrow$ labels spins. 
$n_{\vect r \varepsilon \sigma} = d_{\vect r \varepsilon \sigma}^\dag d_{\vect r \varepsilon \sigma }$ 
is the electron density for orbital-$\varepsilon$ and spin-$\sigma$, 
$n_{\vect r \varepsilon}=\sum_\sigma n_{\vect r \varepsilon \sigma}$ is the electron density for orbital-$\varepsilon$, 
and $n_{\vect r} = \sum_\varepsilon n_{\vect r \varepsilon}$ is the total electron density at site-$\vect r$. 
$U$ is on-site intraorbital repulsion, $U'$ is on-site interorbital repulsion, and $V$ is nearest-neighbor repulsion.

Now we 1/3-fill the 8th band, and try to find out the ground state.
We project the interaction $H_{I}$
into the 8th band, and study the effective Hamiltonian in the partially filled band (the 8th band). 
This treatment does not include the band-mixing due to the interaction so it is not exact. 
But it is a reliable treatment when interaction $U$, $U'$, $V$ are weak compared with the band energy spacing. 
In addition, band-mixing can be included in future study by perturbation treatment.

\begin{eqnarray}
H_{eff}=\sum_{\vect k}E_{8}(k)\psi_{\vect k}^{\dagger}\psi_{\vect k}+
\frac{1}{N_{uc}}\sum_{\vect k_1 \vect k_2 \vect k_3} u(\vect k_1, \vect k_2, \vect k_{3})
\psi_{\vect k_1}^{\dagger}\psi_{\vect k_2}^{\dagger}\psi_{\vect k_3}\psi_{\vect k_1+ \vect k_2-\vect k_3}, 
\label{eq:Heff}
\end{eqnarray}
where $E_8(k)$ is the kinetic energy of the 8th band, and the interaction $u$ is nothing but $H_{I}$ projected into the 8th band. 
$N_{uc}=N_{x}\cdot N_{y}$ is the total number of unit cells. 
$1/N_{uc}$ is the correct normalization factor.

$H_{eff}$ is exact diagonalized for $4\times6$ ($N_{x}\times N_{y}$,
these are the number of unit cells along $\vect a_{1}$ and $\vect a_{2}$ directions
on honeycomb lattice) unit-cell system with the periodic boundary condition,
with $N_{e}=8$ electrons. Because $H_{eff}$ respects the total momentum,
we can diagonalize $H_{eff}$ with in each center-of-mass (COM) momentum sector. 
In Supplementary Figure~S2, 
we show the ground state and the first excited state energies 
as a function of the center-of-mass momentum. 
We choose $U=U'=t,V=0.5t$. The momentum $k_{x},k_{y}$ are shown as integers. 
For example, $(k_{x},k_{y})=(2,3)$ really means $(k_{x},k_{y})=(2\cdot2\pi/N_{x},3\cdot2\pi/N_{y})$.

A three-fold degenerate ground state manifold (GSM) is observed,  
which is separated with the other states by a clear energy gap $\sim0.1$. 
Note that they are not exactly degenerate on a finite system. 
But, their energy difference should fall off exponentially as the system size increases.
These three ground states are at momentum $(0,0),(0,2),(0,4)$ which is expected. 
The reason is that the different ground states can be viewed as a result of twisted boundary condition $0\rightarrow2\pi$. 
If we twist the boundary condition along the $y$ direction $0\rightarrow2\pi$, the momentum of each electron is shifted: 
$k_{y}\rightarrow k_{y}+\frac{2\pi}{6}$. 
So for $8$ electrons, the center of mass momentum shifts $k_{y}\rightarrow k_{y}+8\frac{2\pi}{6}=k_{y}+\frac{2\pi}{3}$.
This twist will drive ground state 1 with center of mass momentum (0,0) to ground state 2 with COM $\vect k=(0,2)$. 
And it also drives ground state 2 into ground state 3.

To confirm that this 3-fold degenerate ground state is really a fractional quantum Hall (FQH) state instead of states such as CDW, 
we computed the Chern number by twist boundary condition. 
The details of the method are described in Ref.~61. 
Here we discretize the boundary phase unit cell into a $10 \times 10$ and $20 \times 20$ meshes. 
And the Chern numbers for the three ground states are found to be independent of which mesh to use up to the fourth digit:
$C_{1}=0.3344,C_{2}=0.3311,C_{3}=0.3344$. 
These values slightly deviate from $C=1/3$ in the thermodynamic limit, which is expected for a small system. 
The sum of the three Chern numbers is found to be exactly $1$. 
This explicitly shows that we are in the $\nu=1/3$ FQH phase.


\newpage
\textbf{Supplementary References}
\begin{itemize}


\item[61] Sheng, D. N., Gu, Z.-C., Sun, K., \& Sheng, L. 
Fractional quantum Hall effect in the absence of Landau levels. 
\emph{Nat. Commun.} {\bf 2}:389 doi: 10.1038/ncomms1380 (2011). 

\end{itemize}


\begin{thebibliography}{10}

\bibitem{QHE}
See, e.g., \textit{The Quantum Hall Effect}, edited by R. E. Prange
and S. M. Girvin (Springer-Verlag, Berlin, 1987).

\bibitem{thouless1982}
Thouless, D.J., Kohmoto, M., Nightingale, M.P. \& den Nijs, M. Quantized Hall Conductance in a Two-Dimensional Periodic Potential. 
\textit{Phys. Rev. Lett.} \textbf{49}, 405 (1982).

\bibitem{haldane1988}
Haldane, F.D.M. Model for a Quantum Hall Effect without Landau Levels: Condensed-Matter Realization of the ``Parity Anomaly.'' 
\textit{Phys. Rev. Lett.} \textbf{61}, 2015 (1988).

\expandafter\ifx\csname url\endcsname\relax
  \def\url#1{\texttt{#1}}\fi
\expandafter\ifx\csname urlprefix\endcsname\relax\def\urlprefix{URL }\fi
\providecommand{\bibinfo}[2]{#2}
\providecommand{\eprint}[2][]{\url{#2}}

\bibitem{kane2005a}
\bibinfo{author}{Kane, C.~L.} \& \bibinfo{author}{Mele, E.~J.}
\newblock \bibinfo{title}{$z_{2}$ topological order and the quantum spin {H}all
  effect}.
\newblock \emph{\bibinfo{journal}{Phys. Rev. Lett.}}
  \textbf{\bibinfo{volume}{95}}, \bibinfo{pages}{146802}
  (\bibinfo{year}{2005}).

\bibitem{bernevig2006}
\bibinfo{author}{Bernevig, B.~A.}, \bibinfo{author}{Hughes, T.~L.} \&
  \bibinfo{author}{Zhang, S.-C.}
\newblock \bibinfo{title}{Quantum spin {H}all effect and topological phase
  transition in {HgTe} quantum wells}.
\newblock \emph{\bibinfo{journal}{Science}} \textbf{\bibinfo{volume}{314}},
  \bibinfo{pages}{1757--1761} (\bibinfo{year}{2006}).

\bibitem{moore2007}
\bibinfo{author}{Moore, J.~E.} \& \bibinfo{author}{Balents, L.}
\newblock \bibinfo{title}{Topological invariants of time-reversal-invariant
  band structures}.
\newblock \emph{\bibinfo{journal}{Phys. Rev. B}} \textbf{\bibinfo{volume}{75}},
  \bibinfo{pages}{121306} (\bibinfo{year}{2007}).

\bibitem{fu2007}
\bibinfo{author}{Fu, L.}, \bibinfo{author}{Kane, C.~L.} \&
  \bibinfo{author}{Mele, E.~J.}
\newblock \bibinfo{title}{Topological insulators in three dimensions}.
\newblock \emph{\bibinfo{journal}{Phys. Rev. Lett.}}
  \textbf{\bibinfo{volume}{98}}, \bibinfo{pages}{106803}
  (\bibinfo{year}{2007}).

\bibitem{konig2007}
\bibinfo{author}{K{\"o}nig, M.} \emph{et~al.}
\newblock \bibinfo{title}{Quantum spin {H}all insulator state in {HgTe} quantum
  wells}.
\newblock \emph{\bibinfo{journal}{Science}} \textbf{\bibinfo{volume}{318}},
  \bibinfo{pages}{766--770} (\bibinfo{year}{2007}).

\bibitem{hsieh2008}
\bibinfo{author}{Hsieh, D.} \emph{et~al.}
\newblock \bibinfo{title}{A topological dirac insulator in a quantum spin
  {H}all phase}.
\newblock \emph{\bibinfo{journal}{Nature}} \textbf{\bibinfo{volume}{452}},
  \bibinfo{pages}{970--974} (\bibinfo{year}{2008}).

\bibitem{xia2009}
\bibinfo{author}{Xia, Y.} \emph{et~al.}
\newblock \bibinfo{title}{Observation of a large-gap topological-insulator
  class with a single {D}irac cone on the surface}.
\newblock \emph{\bibinfo{journal}{Nature Phys.}} \textbf{\bibinfo{volume}{5}},
  \bibinfo{pages}{398--402} (\bibinfo{year}{2009}).

\bibitem{qi2008}
\bibinfo{author}{Qi, X.-L.}, \bibinfo{author}{Hughes, T.~L.} \&
  \bibinfo{author}{Zhang, S.-C.}
\newblock \bibinfo{title}{Topological field theory of time-reversal invariant
  insulators}.
\newblock \emph{\bibinfo{journal}{Phys. Rev. B}} \textbf{\bibinfo{volume}{78}},
  \bibinfo{pages}{195424} (\bibinfo{year}{2008}).

\bibitem{fu2008}
\bibinfo{author}{Fu, L.} \& \bibinfo{author}{Kane, C.~L.}
\newblock \bibinfo{title}{Superconducting proximity effect and {M}ajorana
  fermions at the surface of a topological insulator}.
\newblock \emph{\bibinfo{journal}{Phys. Rev. Lett.}}
  \textbf{\bibinfo{volume}{100}}, \bibinfo{pages}{096407}
  (\bibinfo{year}{2008}).

\bibitem{izumi2001}
\bibinfo{author}{Izumi, M.} \emph{et~al.}
\newblock \bibinfo{title}{Perovskite superlattices as tailored materials of
  correlated electrons}.
\newblock \emph{\bibinfo{journal}{Mat. Sci. Eng. B}}
  \textbf{\bibinfo{volume}{84}}, \bibinfo{pages}{53--57}
  (\bibinfo{year}{2001}).

\bibitem{ohtomo2002}
\bibinfo{author}{Ohtomo, A.}, \bibinfo{author}{Muller, D.~A.},
  \bibinfo{author}{Grazul, J.~L.} \& \bibinfo{author}{Hwang, H.~Y.}
\newblock \bibinfo{title}{Artificial charge-modulation in atomic-scale
  perovskite titanate superlattices}.
\newblock \emph{\bibinfo{journal}{Nature}} \textbf{\bibinfo{volume}{419}},
  \bibinfo{pages}{378--380} (\bibinfo{year}{2002}).

\bibitem{ohtomo2004}
\bibinfo{author}{Ohtomo, A.} \& \bibinfo{author}{Hwang, H.~Y.}
\newblock \bibinfo{title}{A high-mobility electron gas at the
  {La}{Al}{O}$_3$/{Sr}{Ti}{O}$_3$ heterointerface}.
\newblock \emph{\bibinfo{journal}{Nature}} \textbf{\bibinfo{volume}{427}},
  \bibinfo{pages}{423--426} (\bibinfo{year}{2004}).

\bibitem{mannhart2010}
\bibinfo{author}{Mannhart, J.} \& \bibinfo{author}{Schlom, D.~G.}
\newblock \bibinfo{title}{Oxide interfaces---an opportunity for electronics}.
\newblock \emph{\bibinfo{journal}{Science}} \textbf{\bibinfo{volume}{327}},
  \bibinfo{pages}{1607--1611} (\bibinfo{year}{2010}).

\bibitem{shitade2009}
Shitade, A. \emph{et al.} 
\newblock Quantum Spin Hall Effect in a Transition Metal Oxide Na$_{2}$IrO$_{3}$. 
\newblock \emph{Phys. Rev. Lett.} \textbf{102}, 256403 (2009).

\bibitem{pesin2010}
Pesin, D. \& Balents, L. 
\newblock Mott physics and band topology in materials with strong spin-orbit interaction. 
\newblock \emph{Nature Phys.} \textbf{6}, 376 -- 381 (2010). 

\bibitem{fu2006}
\bibinfo{author}{Fu, L.} \& \bibinfo{author}{Kane, C.~L.}
\newblock \bibinfo{title}{Time reversal polarization and a $z_{2}$ adiabatic
  spin pump}.
\newblock \emph{\bibinfo{journal}{Phys. Rev. B}} \textbf{\bibinfo{volume}{74}},
  \bibinfo{pages}{195312} (\bibinfo{year}{2006}).

\bibitem{min2006}
\bibinfo{author}{Min, H.} \emph{et~al.}
\newblock \bibinfo{title}{Intrinsic and {R}ashba spin-orbit interactions in
  graphene sheets}.
\newblock \emph{\bibinfo{journal}{Phys. Rev. B}} \textbf{\bibinfo{volume}{74}},
  \bibinfo{pages}{165310} (\bibinfo{year}{2006}).

\bibitem{yao2007}
\bibinfo{author}{Yao, Y.}, \bibinfo{author}{Ye, F.}, \bibinfo{author}{Qi,
  X.-L.}, \bibinfo{author}{Zhang, S.-C.} \& \bibinfo{author}{Fang, Z.}
\newblock \bibinfo{title}{Spin-orbit gap of graphene: First-principles
  calculations}.
\newblock \emph{\bibinfo{journal}{Phys. Rev. B}} \textbf{\bibinfo{volume}{75}},
  \bibinfo{pages}{041401} (\bibinfo{year}{2007}).

\bibitem{vallin1970}
Vallin, J. T.
Dynamic Jahn-Teller Effect in the Orbital $^2E$ State of Fe$^{2+}$ in CdTe.
\emph{Phys. Rev. B} {\bf 2}, 2390 (1970).

\bibitem{chen2009}
Chen, G., Balents, L., \& Schnyder, A. P. 
Spin-Orbit Singlet and Quantum Critical Point on the Diamond Lattice: FeSc$_2$S$_4$.
\emph{Phys. Rev. Lett.} {\bf 102}, 096406 (2009).

\bibitem{jahn1937}
Jahn, H. A. \& Teller, E. 
\newblock Stability of Polyatomic Molecules in Degenerate Electronic States. I. Orbital Degeneracy.
\newblock \emph{Proc. Roy. Soc. London}
  \textbf{161}, 220--235 (1937).


\bibitem{imada1998}
Imada, M., Fujimori, A. \& Tokura, Y.
\newblock Metal-insulator transitions.
\newblock \emph{Rev. Mod. Phys.} \textbf{70}, 1039--1263 (1998).


\bibitem{yoshimatsu2010}
Yoshimatsu, K. \emph{et al.} 
\newblock Dimensional-Crossover-Driven Metal-Insulator Transition in SrVO$_3$ Ultrathin Films. 
\newblock \emph{Phys. Rev. Lett.} \textbf{104}, 147601 (2010). 


\bibitem{sugiyama1999}Sugiyama, T. \& Tsuda, N. Electrical and Magnetic Properties of Ca$_{1-x}$La$_x$RuO$_3$. 
\emph{J. Phys. Soc. Jpn.} {\bf 68}, 3980--3987 (1999).  



\bibitem{yamaura2001}Yamaura, K. \& Takayama-Muromachi, E. 
Enhanced paramagnetism of the 4$d$ itinerant electrons in the rhodium oxide perovskite SrRhO$_3$. 
\emph{Phys. Rev. B }{\bf 64}, 224424 (2001). 

\bibitem{europeanpatent}MULTILAYER ELECTRODES FOR FERROELECTRIC DEVICES, European Patent EP0636271. 

\bibitem{cao2007}Cao G., Durairaj, V.,. Chikara, S., DeLong, L.E., Parkin, S., \& Schlottmann, P.
Non-Fermi-liquid behavior in nearly ferromagnetic SrIrO$_3$ single crystals. 
\emph{Phys. Rev. B} {\bf 76}, 100402(R) (2007). 

\bibitem{moon2008}Moon, S. J., Jin, H., Kim, K. W.,  Choi, W. S. Lee, Y. S., Yu, J., Cao, G., Sumi, A., Funakubo, H., Bernhard, C., \& Noh T. W. 
Dimensionality-Controlled Insulator-Metal Transition and Correlated Metallic State in 5$d$ Transition Metal Oxides 
Sr$_{n+1}$Ir$_n$O$_{3n+1}$ ($n=1,2,$ and $\infty$). 
\emph{Phys. Rev. Lett.} {\bf 101}, 226402 (2008). 

\bibitem{sumi2005}Sumi, A., Kim, Y. K., Oshima, N., Akiyama, K., Saito, K. \& Funakubo, H.  
MOCVD growth of epitaxial SrIrO$_3$ films on (111)SrTiO$_3$ substrates. 
\emph{Thin Solid Films} {\bf 486}, 182--185 (2005).

\bibitem{bacalis1988}Bacalis, N.C. Band Structure and Electron-Phonon Interaction of LaAgO$_3$. 
\emph{J. Superconductivity} {\bf 1}, 175--180 (1988).

\bibitem{curuswany1980}Guruswany, V.,  Keillor, P., Campbell, G. L., \& Bockris, J. O'M. 
The photoelectrochemical response of the lanthanides of chromium, rhodium, vanadium and gold on a titanium base. 
\emph{Solar Energy Materials} {\bf 4}, 11--30 (1980).

\bibitem{ralle1993}Ralle, M. \& Jansen, M.  
Synthesis and Crystal Structure Determination of LaAuO$_3$. 
\emph{J. Solid State Chem.} {\bf 105}, 378--384 (1993).


\bibitem{Chakraverty2010a}Chakraverty, S., Ohtomo, A., Okude, M., Ueno, K., \& Kawasaki, M.
Epitaxial Structure of (001)- and (111)-Oriented Perovskite Ferrate Films Grown by Pulsed-Laser Deposition. 
\emph{Cryst. Growth Des.} {\bf 10}, 1725E1729 (2010). 

\bibitem{gray2010}
Gray, B., Lee, H.-Y., Liu, J., Chakhalian, J., \& Freeland, J. W.
Local electronic and magnetic studies of an artificial La$_2$FeCrO$_6$ double perovskite. 
\emph{Appl. Phys. Lett.} {\bf 97}, 013105 (2010). 






\bibitem{nagaosa2010}
Nagaosa, N. \emph{et al.} 
\newblock Anomalous Hall effect. 
\newblock \emph{Rev. Mod. Phys.} \textbf{82}, 1539--1592 (2010). 

\bibitem{tang2010}Tang, E., Mei, J.-W., \& Wen, X.-G. 
High-Temperature Fractional Quantum Hall States.
\emph{Phys. Rev. Lett.} {\bf 106}, 236802 (2011).

\bibitem{sun2010}
Sun, K., Gu, Z., Katsura, H., \& Das Sarma, S.
Nearly Flatbands with Nontrivial Topology. 
\emph{Phys. Rev. Lett.} {\bf 106}, 236803 (2011).

\bibitem{neupert2010}Neupert, T., Santos, L., Chamon, C., \& Mudry, C.  
Fractional Quantum Hall States at Zero Magnetic Field.  
\emph{Phys. Rev. Lett.} {\bf 106}, 236804 (2011).

\bibitem{venderbos2011}Venderbos, J. W. F., Daghofer, M., and van den Brink, J.
Narrowing of Topological Bands due to Electronic Orbital Degrees of Freedom.
\emph{Phys. Rev. Lett.} {\bf 107}, 116401 (2011) 

\bibitem{wang2011}Wang, Y.-F., Gu, Z.-C., Gong, C.-D., \& Sheng, D. N. 
Fractional Quantum Hall Effect of Hard-Core Bosons in Topological Flat Bands.
\emph{Phys. Rev. Lett.} {\bf 107}, 146803 (2011).


\bibitem{neupert2011}Neupert, T., Santos, L., Ryu, S., Chamon, C., \& Mudry, C.  
Fractional topological liquids with time-reversal symmetry and their lattice realization.  
Phys. Rev. B {\bf 84}, 165107 (2011).

\bibitem{bonderson2006}
Bonderson, P., Kitaev, A., \& Shtengel, K.
Detecting Non-Abelian Statistics in the $\nu=5/2$ Fractional Quantum Hall State. 
\emph{Phys. Rev. Lett.} {\bf 96}, 016803 (2006). 

\bibitem{nayak2008}
Nayak, C., Simon, S. H., Stern, A., Freedman, M., \& Das Sarma, S. 
Non-Abelian Anyons and Topological Quantum Computation.
\emph{Rev. Mod. Phys.} {\bf 80}, 1083--1159 (2008). 

\bibitem{moore1991}
Moore, G. \& Read, N.
Nonabelions in the fractional quantum Hall effect. 
\emph{Nucl. Phys. B} {\bf 360}, 362--396 (1991).

\bibitem{levin2007}
Levin, M., Halperin, B. I., \& Rosenow, B.
Particle-hole symmetry and the Pfaffian state. 
\emph{Phys. Rev. Lett.} {\bf 99}, 236806 (2007). 

\bibitem{lee2007}
Lee, S.-S., Ryu, S., Nayak, C., \& Fisher, M. P. A. 
Particle-Hole Symmetry and the $\nu={5/2}$ Quantum Hall State.
\emph{Phys. Rev. Lett.} {\bf 99}, 236807 (2007). 



\bibitem{murakami2006}Murakami, S. 
Quantum Spin Hall Effect and Enhanced Magnetic Response by Spin-Orbit Coupling. 
\emph{Phys. Rev. Lett.} {\bf 97}, 236805 (2006). 

\bibitem{wangF2011}Wang, F. \& Ran, Y. 
Nearly flat band with Chern number $C=2$ on the dice lattice. 
Preprint at $<$http://arxiv.org/abs/1109.3435$>$ (2011). 


\bibitem{meng2010}Meng, Z. Y., Lang, T. C., Wessel, S., Assaad, F. F., \& Muramatsu, A.
Quantum spin liquid emerging in two-dimensional correlated Dirac fermions.
\emph{Nature} {\bf 464}, 847--852 (2010).

\bibitem{ruegg2011}
R{\"u}egg, A. \& Fiete, G. A. 
Topological insulators from complex orbital order in transition-metal oxides heterostructures. 
\emph{Phys. Rev. B} {\bf 84},  201103(R) (2011).

\bibitem{yang2011}Yang, K.-Y., Zhu, W., Xiao, D., Okamoto, S., Wang, Z., \& Ran, Y.
Possible interaction-driven topological phases in (111) bilayers of LaNiO$_3$.
\emph{Phys. Rev. B} {\bf 84},  201104(R) (2011).

\bibitem{slater54} Slater, J. C. \& Koster, G. F. 
Simplified LCAO Method for the Periodic Potential Problem. 
\emph{Phys. Rev.} {\bf 94}, 1498--1524 (1954). 

\bibitem{kresse1999}Kresse, G. \& Joubert, D. From ultrasoft pseudopotentials to the projector augmented-wave method. 
\emph{Phys. Rev. B} {\bf 59}, 1758--1775 (1999).
\bibitem{perdew1996}Perdew, J. P., Burke, K. \& Ernzerhof, M. Generalized gradient approximation made simple. 
\emph{Phys. Rev. Lett.} {\bf 77}, 3865--3868 (1996).
\bibitem{kresse1996}Kresse, G. \& Furthm{\"u}ller, J. 
Efficient iterative schemes for Ab Initio total-energy calculations using a plane-wave basis set. 
\emph{Phys. Rev. B} {\bf 54}, 11169--11186 (1996).
\bibitem{berkstresser1991}Berkstresser, G. W., Valentino, A. J. \& Brandle, C. D. Growth of single crystals of lanthanum aluminate. 
\emph{J. Cryst. Growth} {\bf 109}, 467E-471 (1991).

\bibitem{hellwege81}
Hellwege, K.-H. \& Hellwege, A. M. (eds) 
\emph{Landolt-B{\"o}rnstein: Numerical Data and Functional Relationships in Science and Technology} New Series, 
Group III, Vol. 16a, 59--64 (Springer, Berlin, 1981).



\end{thebibliography}
\end{document}